# Bright, high-repetition-rate water window soft X-ray source enabled by nonlinear pulse self-compression in an antiresonant hollow-core fibre

M. Gebhardt,[1,2,*] T. Heuermann,[1,2] R. Klas,[1,2] C. Liu,[1,2] A. Kirsche,[1,2] M. Lenski,[1] Z. Wang,[1] C. Gaida,[1,+] J. E. Antonio-Lopez,[3] A. Schülzgen,[3] R. Amezcua-Correa,[3] J. Rothhardt,[1,2,4] and J. Limpert,[1,2,4]

[1]Institute of Applied Physics, Abbe Center of Photonics, Friedrich-Schiller-Universität Jena, Albert-Einstein-Str. 15, 07745 Jena, Germany

[2]Helmholtz-Institute Jena, Fröbelstieg 3, 07743 Jena, Germany

[3]CREOL, College of Optics and Photonics, University of Central Florida, Orlando, Florida 32816, USA

[4]Fraunhofer Institute for Applied Optics and Precision Engineering, Albert-Einstein-Str. 7, 07745 Jena, Germany

[+]now with Active Fiber Systems GmbH, Ernst-Ruska-Ring 17, 07745 Jena, Germany

*Corresponding author: martin.gebhardt@uni-jena.de

**Bright, coherent soft X-ray (SXR) radiation is essential to a variety of applications in fundamental research and life sciences. To date, a high photon flux in this spectral region can only be delivered by synchrotrons, free electron lasers or high-order harmonic generation (HHG) sources, which are driven by kHz-class repetition rate lasers with very high peak powers. Here, we establish a novel route towards powerful and easy-to-use SXR sources by presenting a compact experiment in which nonlinear pulse self-compression to the few-cycle regime is combined with phase-matched HHG in a single, helium-filled antiresonant hollow-core fibre (ARHCF). This enables the first 100 kHz-class repetition rate, table-top SXR source that delivers an application-relevant flux of $2.8\times10^6$ photons $s^{-1}$ $eV^{-1}$ around 300 eV. The fibre integration of temporal pulse self-compression (leading to the formation of the necessary strong-field waveforms) and pressure-controlled phase matching will allow compact, high-repetition-rate laser technology, including commercially available systems, to drive simple and cost-effective, coherent high-flux SXR sources.**

Laser-driven SXR sources based on HHG[1] are known for their table-top dimensions, excellent spatial coherence and ultrashort pulse durations[2], which make them attractive tools for advanced spectroscopy[3,4]. Additionally, they are expected to enable area-wide evolution of lens-less imaging in the water window[5], and they hold great promise for the production of isolated attosecond pulses shorter than the atomic unit of time[6,7].

In the past decade, water window HHG has mostly been achieved with the help of optical parametric amplifiers operating at a wavelength of approximately 2 μm[4,8–11]. This driving wavelength is identified as a "sweet spot" for pushing the phase-matched harmonic energy ($\sim\lambda^{1.4-1.7}$)[12] beyond the carbon K-edge, while the single-atom response ($\sim\lambda^{-(5-6)}$)[13] is still reasonable and can partially be compensated for by high phase matching pressures[14]. To date, the reported generated photon flux around 300 eV is as high as $1\times10^9$ photons $s^{-1}$ $eV^{-1}$ at a 1 kHz repetition rate based on HHG in a gas-filled capillary[4]. To achieve high flux levels, the typical experimental conditions require >40 GW peak power (Supplement), which implies the generation and handling of multi-mJ energy[4,9], or few-cycle pulses[11,15]. These experimental constraints alter the fundamentally desired user-friendly and straightforward nature of laser-based SXR sources and are responsible for the fact that subsequent work on applications is often closely related to source development[4,11]. Additionally, techniques such as coincidence detection[16] and space-charge-reduced photoelectron spectroscopy[17] require repetition rates >1 kHz, where the generation and handling of high-peak-power pulses become increasingly challenging, and the reported 300 eV flux is only approximately $4\times10^4$ photons $s^{-1}$ $eV^{-1}$, based on HHG in a free-space gas target[9,15]. It is therefore highly desired to enable compact, high-repetition-rate laser sources (including turnkey, commercial systems) to directly generate high-flux SXR high-order harmonics.

In this work, we demonstrate an approach to SXR HHG, which is based on power-scalable laser emission from a 98 kHz repetition rate thulium-doped fibre laser. It combines nonlinear self-compression of the driving pulses[18] with waveguide HHG[19] within a single, gas-filled ARHCF. In contrast to previous demonstrations[4,8–11,15], the fibre-integrated HHG experiments require only GW-level peak power, multi-cycle laser pulses. This is a result of the carefully controlled intensity enhancement during pulse compression. While so-called all-fibre, table-top SXR sources, driven by mid-infrared lasers, have been briefly envisaged theoretically[20], this is the first time that this scheme is experimentally realized and analysed in depth. This is fundamentally enabled by advances in the development of ARHCFs[21], which have already enabled HHG in the extreme ultraviolet[22], and represents a significant step in increasing the availability and performance of high-repetition-rate sources for SXR or attosecond science.

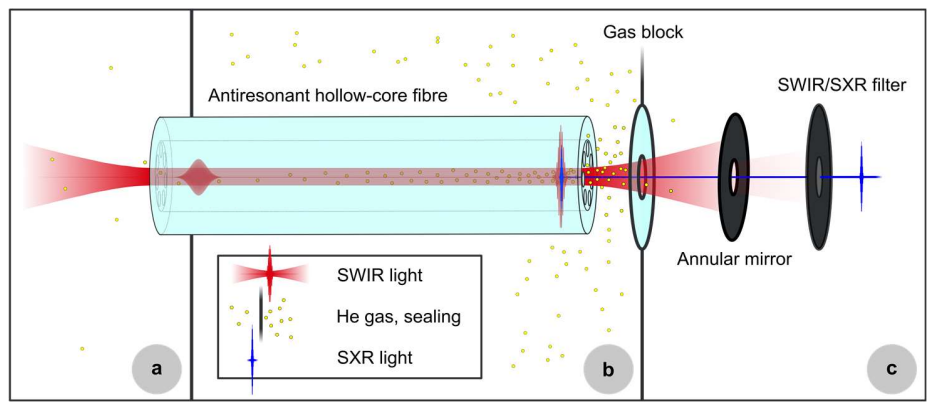

**Fig. 1 Nonlinear pulse self-compression and HHG setup. a,** Input coupling within helium gas at atmospheric pressure. **b,** The ARHCF (core diameter: 84.5 µm, length: 120 cm) is differentially pumped with the output facet located in a high-pressure chamber. Near the fibre output, the high-pressure region is separated from the following vacuum chamber by a 1 mm thick sapphire window that has a conical hole with diameters of 100 µm to 250 µm. This gas block prevents most of the helium from reaching the following low-pressure region, while the light emerging from the fibre passes through it. To achieve this, the ARHCF output (located approximately 2 mm from the gas block) is movable in the horizontal and vertical directions (only the optimal positioning is shown for clarity). **c,** The final vacuum chamber contains an annular mirror, a thin metal filter to separate the SWIR light from the generated high-order harmonics, and a flat-field imaging grating spectrometer (not shown).

Our experiments are enabled by a thulium-doped fibre chirped-pulse amplification system that provides 100 fs pulses with energies up to 450 µJ at a central wavelength of 1910 nm (Methods, Supplement). To prevent absorption from water vapor or thermal failure of the fibre tip[23], mode coupling to the ARHCF is performed within helium gas at 1 bar pressure (Fig. 1a). Through its output end, the fibre can be filled with helium at a pressure of up to 20 bar. Even at this pressure, the zero-dispersion wavelength is below the central wavelength of the laser pulses (Methods). Hence, these pulses self-compress as they undergo self-phase modulation while propagating to the output of the waveguide. For appropriately chosen input pulse energy and output pressure, the enhanced electric field strength close to the fibre end is high enough to directly drive HHG within the ARHCF (Fig. 1b). We note that the typical phase-matched photon energy cut-off achieved with short-wavelength infrared (SWIR) driving wavelengths is in the SXR regime[12]. Because of the good transparency of helium in this spectral region, the high-photon-energy portion of the generated harmonics is well transmitted to the characterization (Fig. 1c) or subsequent experiments.

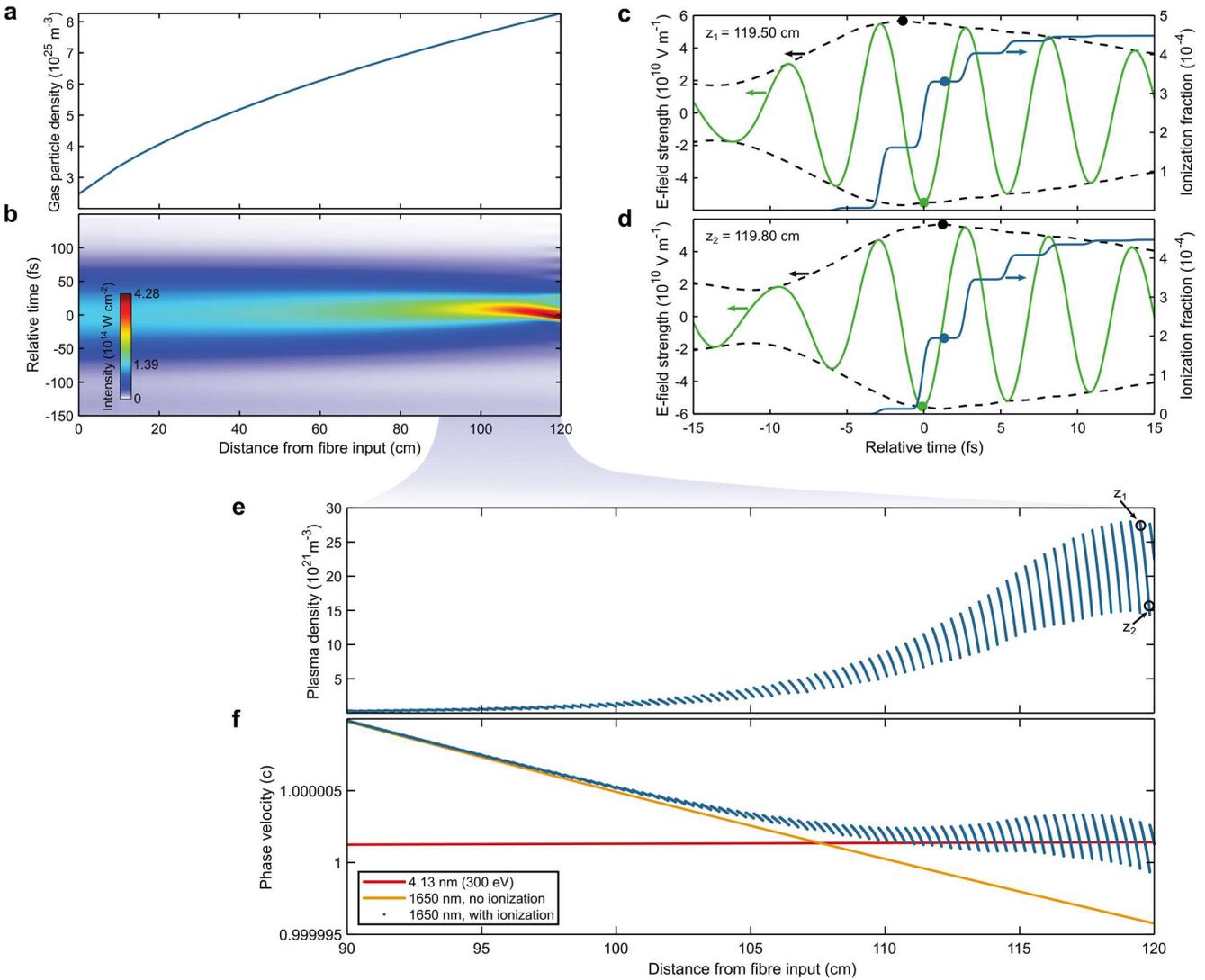

**Fig. 2 Nonlinear pulse self-compression and simultaneous phase matching of SWIR and SXR light. a,** Helium gas particle density along the fibre for pressures of 1 bar at the input and 3.345 bar at the output. **b,** Evolution of the simulated time-dependent on-axis intensity within the fibre. The reference frame velocity is the group velocity at 1910 nm. **c,** Simulated time-dependent electric field strength (green line), envelope (black dashed line) and ionization fraction (blue line) at a distance $z_1$=119.50 cm from the fibre input. Solid points highlight the field peak values/positions and the ionization fraction after the global peak of the electric field. The relative time axis is centred at the electric field maximum at $z_1$. **d,** same as **c**, but at $z_2$=119.80 cm from the fibre input. The reference frame velocity is the phase velocity at 1650 nm. **e,** Simulated on-axis plasma density after the global peak of the electric field. **f,** Comparison of the SWIR and SXR phase velocities vs. distance from the fibre input.

A numerical analysis of the pulse self-compression (Methods) and phase matching to the SXR light is presented in Fig. 2. The input pulse characteristics are chosen to represent the experimental conditions (Supplement). The steady-state gas particle density within the ARHCF (Fig. 2a) is retrieved from gas flow modelling (Methods), with a pressure of 3.345 bar at the fibre output. Fig. 2b shows the evolution of the time-dependent on-axis intensity inside the fibre for a launched pulse energy of 265 µJ. The nonlinear dynamics cause the pulses to self-compress from an initial FWHM duration of 100 fs to below 20 fs (3.5 optical cycles) as they reach a position just before the fibre output. At this point, the peak intensity is increased to $4.28\times10^{14}$ W cm$^{-2}$, which causes ionization of the gas, as seen from the acceleration of the pulse (Fig. 2b)[24]. To observe a significant growth of harmonic radiation, it is necessary to fulfil the phase matching conditions[1] for at least one half-cycle within the SWIR pulse. Due to the significant field strength dependence of the atomic response[25], phase matching close to the peak of the driving pulse is preferred. This is also favourable for macroscopic harmonic signal growth from few-cycle driving pulses because it reduces the variation in the intrinsic HHG phase with propagation distance[14]. Consequently, we evaluate the ionization levels directly after the peak of the pulse. For example, Fig. 2c shows the simulated time-dependent electric field and ionization fraction at a distance $z_1$=119.50 cm. This is compared to the situation at $z_2$=119.80 cm (Fig. 2d). It becomes apparent that the evolution of the on-axis plasma density directly after the global electric field maximum (Fig. 2e) exhibits a modulation due to the temporal walk-off between the carrier and its envelope. This walk-off is a direct consequence of the waveguide dispersion, similar to the Gouy phase shift in a tight focusing geometry. It can be furthermore derived from the self-compressed electric fields that the instantaneous wavelength of the most intense cycle is approximately 1650 nm. This is not surprising, as ionization causes a blueshift of the spectrum[24]. Finally, the above derived quantities are used to calculate the phase velocities for the centre of the HHG driving pulse and a desired high-order harmonic wavelength. It can be seen in Fig. 2f that the chosen output pressure not only is crucial for the intensity enhancement achieved through pulse self-compression but also, most importantly, counteracts the ionization-induced acceleration of the phase velocity of the driving field such that it can travel in-phase with the generated SXR light. While this situation is globally maintained over several centimetres, we find that the abovementioned walk-off effects limit the coherent growth of the harmonic signal from an individual field half-cycle to a few millimetres (Supplement). This is consistent with the steep slope of the individual half-cycle phase velocities (Fig. 2f, blue data points) and requires fine tuning of the experimental conditions for maximum photon flux.

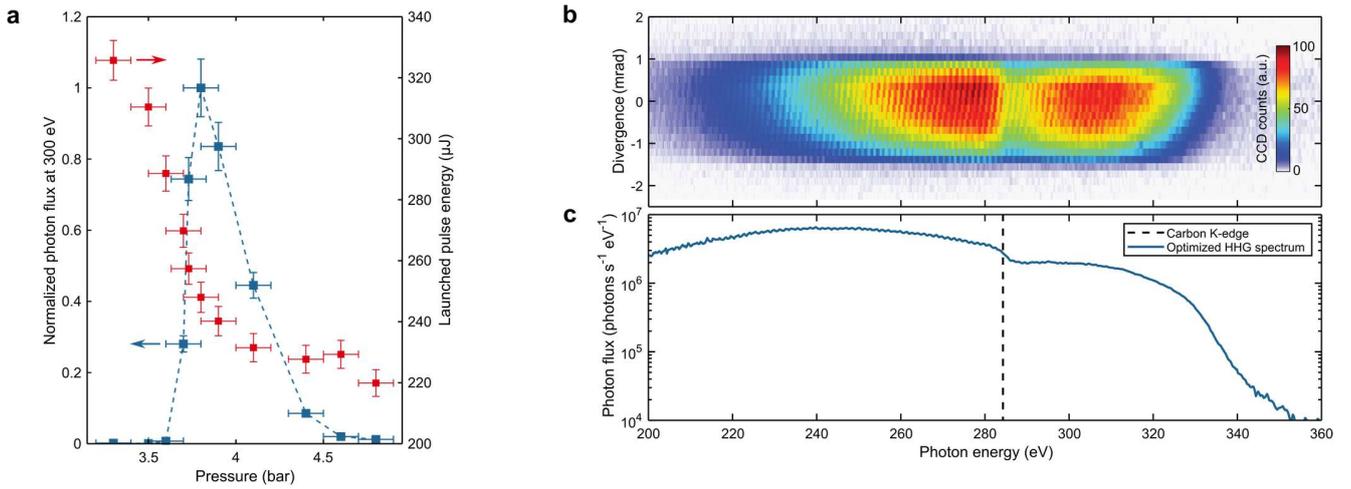

**Fig. 3 High-harmonic generation experimental results. a,** Normalized generated photon flux, evaluated at 300 eV for different helium gas pressures applied to the high-pressure chamber. The individual error bars in the y-direction are based on the standard deviation of 240 consecutive measurements (Supplement) and the uncertainty of the pressure reading (0.1 bar). Measured flux data points are connected by the blue dashed line. The right y-axis refers to the launched pulse energy necessary for obtaining the optimum flux at each pressure setting. **b,** Measured HHG spectrum and divergence under optimized experimental conditions. The clipping in the divergence direction is due to the limited apertures of the SWIR/SXR separation elements. **c,** Evaluation of the generated photon flux not accounting for carbon contaminants, as can be seen from the K-edge absorption feature at a 284 eV photon energy.

The experimental conditions are optimized with the goal of generating a high photon flux around 300 eV. This is done by scanning the gas pressure applied to the high-pressure chamber and subsequently adjusting the launched pulse energy for optimum flux (Fig. 3a). The optimization parameter exhibits a sharp peak around a pressure of 3.8 bar, which agrees reasonably well with the expectations from our numerical analysis (Fig. 2). A drop in flux to $\sim 1/e^2$ times the optimum value is observed for a relative pressure change of 10%. In contrast, the pulse propagation simulations show that the overall nonlinearity of the pulse self-compression and, consequently, the peak electric field strength at the fibre output are easily restored when adjusting the launched pulse energy to a 10% pressure change. Therefore, we conclude that controlling the gas pressure allows control of the phase matching and favouring of the coherent growth of SXR light in the ARHCF. Fig. 3b presents a measured HHG spectrum including the beam divergence, which is represented on the vertical axis. For the signal-optimized experimental conditions, a full-angle divergence of approximately 2 mrad (Methods) and a phase-matched photon energy cut-off around 330 eV are observed. In fact, the simulated, phase-matched half-cycle cut-off is 334 eV (Supplement). In the experiment presented herein, an overall flux of $(2.8\pm1.8)\times10^6$ photons $s^{-1}$ $eV^{-1}$ is generated at 300 eV (Fig. 3c, Methods). Using only one annular mirror and one 200 nm thin aluminium filter, the SWIR suppression is $>10^6$, with $(1.3\pm0.8)\times10^5$ photons $s^{-1}$ $eV^{-1}$ available for experiments. Subsequently, we demonstrate X-ray absorption fine-structure spectroscopy near the carbon K-edge, and we measure an integrated power RMS stability of <5% in the water window over 20 minutes (Supplement). This shows that our table-top, 100 kHz-class repetition rate water window source operates at an application-relevant flux and performance level.

In conclusion, we have demonstrated nonlinear pulse self-compression and SXR HHG at a 98 kHz repetition rate in a single ARHCF. This results in >$10^6$ photons s$^{-1}$ eV$^{-1}$ at 300 eV, directly emitted from the fibre output. This is the first time that such an experiment is reported, and the results show that high-repetition-rate ultrafast lasers that deliver moderate pulse peak powers can be used in a simple, integrated scheme to generate high-order harmonics up to the water window. Because the SXR yield is not absorption-limited and the involved ionization levels are below the critical limit for phase matching, we see significant prospects for advancing the approach described herein in terms of photon flux (Supplement). Furthermore, with an increase in the driving laser wavelength[26], self-compression in ARHCFs will allow direct generation of keV photon energies and extremely short attosecond pulses at a ≥100 kHz repetition rate.

While our work opens up possibilities to study complementary techniques such as quasi-phase matching[22] or driving pulse synthesis[27] within the waveguide, we believe that its results are most interesting for a variety of applications that significantly benefit from compact and easy-to-use high-repetition-rate SXR sources, e.g., coincidence spectroscopy[16] or table-top lens-less imaging of organic samples[5]. Because of the versatility of the pulse self-compression, this approach can be scaled to much higher peak powers[18]. However, we believe that this work represents, first and foremost, a milestone in the development of industrial-grade, laser-driven SXR sources. These sources could use ARHCFs for beam delivery[28], self-compression and HHG in a single apparatus, making them more affordable and available to a much broader community in fundamental and applied sciences with medical applications in reach.

## Methods

### Fibre laser setup

The ultrafast fibre laser system is seeded by a commercial seed source and uses thulium-doped photonic crystal fibres within the first preamplification stages. The pulses are stretched to a duration of approximately 1 ns using a grating-based, Oeffner-type stretcher before further amplification and reduction of the oscillator repetition rate based on an acousto-optic modulator. The pulse train with a 98 kHz repetition rate has an average power of approximately 200 mW prior to the main amplifier of the system, which consists of a thulium-doped large-pitch fibre with a core diameter of 80 µm. The main amplifier is pumped with a commercially available 793 nm diode laser and increases the average power to approximately 50 W. After pulse compression in a grating-based Treacy-type compressor enclosed in a vacuum chamber, the output pulses carry up to 450 µJ of energy and have a duration (FWHM) of 100 fs.

### Dispersion of the ARHCF

The dispersion of the ARHCF is modelled as described in ref.[20]. The fibre used in the experiments described herein has a resonance band up to a wavelength of 1380 nm. This resonance is accounted for by an additional term added to the propagation constant[29], which we use for the pulse propagation simulations. We find the zero-dispersion wavelength (<700 nm in this case) by numerically calculating the root of the group velocity dispersion.

### Calculation of the gas flow and gas density

The gas flow through the ARHCF is modelled as described in[30], assuming a circular tube with a 84.5 µm diameter and a 120 cm length. It can be shown that the continuum component of the flow strongly dominates over the molecular flow component, allowing us to use the well-known description for the pressure and density distribution[31]

$$p(z) = \sqrt{p_0^2 + \frac{z}{L}(p_L^2 - p_0^2)}$$

where $p_L$ and $p_0$ are the fixed pressures at z = L, the fibre output, and at z = 0, the fibre input.

### Simulation of pulse propagation

Numerical simulations of pulse propagation in the fibre are performed based on the unidirectional field propagation equation for the fundamental mode[32]:

$$\partial_z E(z,\omega) = i\left(\beta(\omega) - \frac{\omega}{v}\right) E(z,\omega) + i\frac{\omega^2}{2c^2\epsilon_0\beta(\omega)} P^{\text{NL}}(z,\omega)$$

Here, E is the electric field amplitude in the spectral domain, $\omega$ is the angular frequency, $\beta$ is the fundamental mode propagation constant, $v$ is the reference frame velocity, $c$ is the vacuum speed of light, $\epsilon_0$ is the vacuum permittivity and $P^{\text{NL}}$ is the nonlinear polarization. $P^{\text{NL}}$ is influenced by the Kerr nonlinearity and plasma formation, which we include in the simulations by calculating the Ammosov-Delone-Krainov ionization rates[33].

The fibre is designed for fundamental mode operation, which is consistent with the absence of higher order modes at the fibre output during the experiments. The plasma density in this work is still moderate, from which we deduce that coupling of the power to higher order modes during propagation is not significant. Therefore, pulse propagation is described in the fundamental mode only, for which we find good agreement with the experiments.

**Calculation of phase velocities**

The phase velocity of the high-order harmonics is calculated from the refractive index data available in ref.[34]. For the driving field propagation constant, which gives straightforward access to its phase velocity, we make the approximation described in[1]:

$$\beta(\omega) = \frac{\omega}{c} \cdot \left(1 + \frac{p}{p_0}(1-\eta) \cdot \delta\right) - \frac{2\pi c}{\omega} \cdot \left(\frac{p}{p_0} \eta N_{\text{atm}} r_e + \frac{2.4048^2}{4\pi a^2}\right)$$

where $p$ and $p_0$ are the pressure and the experimental pressure under standard conditions, $\eta$ is the ionization fraction, $\delta$ is the neutral gas dispersion, $N_{\text{atm}}$ is the number density under standard conditions, $r_e$ is the classical electron radius and $a$ is the core radius of the hollow fibre (see Supplement for definition).

**Determination of the launched pulse energy**

In low-power operation, the overall transmission of the 120 cm long ARHCF is approximately 90%. During the HHG experiments, we perform relative power monitoring of the output power, which is used together with the simulation results to determine the experimental launched pulse energy, as shown in Fig. 3a. This gives 86% transmission for the experimental conditions of the optimal HHG signal (see Table S1).

**Characterization of the HHG flux**

The photon flux is estimated from the spectral characterization of the HHG signal. This method is discussed in more detail in ref.[35], where the flux in photons per second, $N_{\text{ph,s}}$, is given as:

$$N_{\text{ph,s}} = \frac{S_{\text{CCD}} \cdot \sigma}{\eta_{\text{QE}} \cdot E_{\text{ph}}^{[\text{bg}]} \cdot \eta_g \cdot t_f \cdot t_{\text{gas}} \cdot t_{\text{exp}}}$$

Here, $S_{\text{CCD}}$ is the signal (counts) measured on the detector, $\sigma$ and $\eta_{\text{QE}}$ are the CCD sensitivity and CCD quantum efficiency, and $E_{\text{ph}}^{[\text{bg}]}$ is the evaluated HHG photon energy in units of a bandgap energy of 3.65 eV. The factor $\sigma/(\eta_{\text{QE}} \cdot E_{\text{ph}}^{[\text{bg}]})$ has been calibrated for the CCD in use by the Physikalisch-Technische Bundesanstalt (PTB). $\eta_g$ is the grating diffraction efficiency (retrieved from measurements using free-electron-laser radiation[36]), $t_f$ is the transmission of the 200 nm aluminium filter used (which is measured experimentally), and $t_\text{gas}$ is the theoretical transmission through the helium gas, consisting of the high-pressure region (see the main text) and the low-pressure spectrometer chamber, for which we determine a propagation length of 1.25 m and a pressure of 0.7 mbar. Finally, $t_\text{exp}$ is the exposure time of the measurement. In addition to these corrections, the clipping of the HHG beam, shown in Fig. 3b, is corrected. For this purpose, a Gaussian profile is fitted to the vertical outline of the signal at 300 eV. This evaluation also gives an estimated full divergence angle of 2 mrad. The combined relative uncertainties of the quantities mentioned above give a worst-case error estimation of ±64% for the absolute flux value.

The visible dip at approximately 284 eV (Fig. 3b and 3c) is associated with the K-shell absorption in carbon and relates to a hydrocarbon contaminant, which is most likely deposited on the spectrometer grating. The contaminant absorption is equivalent to that of a 35 nm thick pure carbon layer.

**Acknowledgements**

The authors would like to thank Prof. Adrian Pfeiffer and Dr. Felix Köttig for fruitful discussions and help with the numerical methods. Furthermore, the authors would like to thank Dr. Steffen Hädrich, Dr. Cesar Jauregui-Misas and Tobias Ulsperger for their help in the early stages of this work.

This work was supported by the European Research Council (ERC) under the European Union's Horizon 2020 research and innovation programme (grant 835306, SALT), the Fraunhofer Cluster of Excellence Advanced Photon Sources (CAPS), the Helmholtz-Institute Jena, the U.S. Army Research Office (grant W911NF1910426) and the U.S. Airforce Office of Scientific Research (grant FA9550-15-10041).

**Author contributions**

J.L., J.R., M.G. and R.K. conceived and planned the experiment. The ultrafast fibre laser and the nonlinear pulse compression stage were built and optimized by M.G., T.H., M.L., Z.W. and C.G. The antiresonant hollow-core fibre was designed and drawn by J.A.-L., A.S. and R.A.-C. The HHG experiments were performed by M.G., T.H., R.K., C.L. and A.K. Simulations and data analysis were performed by M.G.

with support from T.H. and R.K. All authors discussed and contributed to the interpretation of the results and to the writing of the manuscript. J.L. and J.R. supervised the project. J.L., R.A.-C and J.R. acquired funding.

**Competing interests**

The authors declare no competing interests.

Supplementary information accompanies the manuscript on the Light: Science & Applications website (http://www.nature.com/lsa).